\documentclass[10pt,journal]{IEEEtran}

\usepackage{enumitem}
\usepackage{amsmath,graphics,amssymb,epsfig,subfigure,color,cite}

\usepackage{amsthm}
\theoremstyle{plain}
\newtheoremstyle{mystyle}
  {0mm}
  {0mm}
  {}
  {4mm}
  {\bfseries}
  {:}
  { }
  {\thmname{#1}\thmnumber{ #2}\thmnote{ (#3)}}
\theoremstyle{mystyle}

\usepackage{array}
\usepackage{multirow}
\usepackage{algorithmic}
\usepackage{algorithm}
\usepackage{bm}
\usepackage{float}
\usepackage{makecell}
\usepackage[right=0.625in, left=0.625in, top=0.7in, bottom=1.1in]{geometry}
\usepackage[table]{xcolor} 
\usepackage{tablefootnote} 

\newtheorem{thm}{Theorem}
\newtheorem{lem}{Lemma}

\newtheorem{prop}{Proposition}

\floatname{algorithm}{Algorithm}

\newcommand{\vast}{\bBigg@{4.5}}
\newcommand{\Vast}{\bBigg@{7.5}}
\def\blfootnote{\gdef\@thefnmark{}\@footnotetext}

\begin{document}

	\title{Beam-Hopping Pattern Design for Grant-Free Random Access in LEO Satellite Communications}
	
	\author{\author{Seunghyeon Jeon, Seonjung Kim, Gyeongrae Im, and Yo-Seb Jeon
	\thanks{Seunghyeon Jeon, Seonjung Kim, and Yo-Seb Jeon are with the Department of Electrical Engineering, POSTECH, Pohang, Gyeongbuk 37673, South Korea (email: \{seunghyeon.jeon,seonjung.kim,yoseb.jeon\}@postech.ac.kr)} 
    \thanks{Gyeongrae Im is with Electronics and Telecommunications Research Institute (ETRI), Daejeon 34129, South Korea (e-mail: imgrae@etri.re.kr).}
    }

	}
	\vspace{-4mm}
	
	\maketitle
	\vspace{-12mm}

	\begin{abstract} 
    \textcolor{black}{Increasing demand for massive device connectivity in underserved regions drives the development of advanced low Earth orbit (LEO) satellite communication systems.
    Beam-hopping LEO systems without connection establishment provide a promising solution for achieving both demand-aware resource allocation and low access latency.
    This paper investigates beam-hopping pattern design for the grant-free random access systems to dynamically allocate satellite resources according to traffic demands across serving cells.}
    We formulate a binary optimization problem that aims to maximize the minimum successful transmission probability across cells, given limited satellite beam generation capacity.
    To solve this problem, we propose novel beam-hopping design algorithms that alternately enhance the collision avoidance rate and decoding success probability within an alternating optimization framework.
    Specifically, the algorithms employ a bisection method to optimize illumination allocation for each cell based on demand, while using the alternating direction method of multipliers (ADMM) to optimize beam-hopping patterns for maximizing decoding success probability.
    \textcolor{black}{Furthermore, we enhance the ADMM by replacing the strict binary constraint with two equivalent continuous-valued constraints.}
    Simulation results demonstrate the superiority of the proposed algorithms compared to other beam-hopping methods and verify robustness in managing traffic demand imbalance.

	\end{abstract}

    \begin{IEEEkeywords}
    Beam-hopping, low Earth orbit (LEO) satellite, multi-beam  satellite, grant-free random access, alternative optimization
    \end{IEEEkeywords}
	
	
	
	\section{Introduction}\label{Sec:Intro}
    
    
    
    Satellite communication has gained significant attention as a promising solution for providing extensive coverage and ubiquitous connectivity in next-generation wireless networks \cite{Satellite_survey}. Recent advancements in reusable launch systems and decreasing satellite manufacturing costs have made satellite deployment more economically viable, accelerating interest in this technology. 
    While advanced broadband services are available in some regions, vast areas remain underserved due to harsh environmental conditions and high deployment costs. Satellite communication can provide connectivity to unserved regions such as deserts, forests, maritime areas, and polar regions \cite{Satellite_for_5G/6G} and offers reliable communication services during catastrophic situations, including earthquakes, wildfires, and warfare.
    Despite its promise, satellite communication is often constrained by hardware limitations, which restrict both transmission power and onboard processing capabilities \cite{LEO-SAN_challenges}.
    Moreover, traffic demand fluctuates across different regions and time periods, necessitating adaptive resource management to maintain service quality.
    To address these challenges, multi-beam satellites equipped with analog beam steering techniques have been widely adopted \cite{Multi-beam_allocation}.
    These systems enable multiple spot beam generation to serve different regions simultaneously, offering a balance between coverage flexibility and hardware complexity \cite{Spaceborne_multi-beam}.  
    However, such systems typically rely on static beam allocation, which fails to adjust to real-time traffic variations, causing resource inefficiencies.


    \textcolor{black}{Beam-hopping has emerged as a dynamic resource allocation technique that enhances the adaptability of multi-beam satellite systems by flexibly distributing resources according to varying traffic demands \cite{Beam-hopping}.
    Unlike conventional multi-beam systems that provide uniform coverage, beam-hopping selectively illuminates specific cells during each time slot based on patterns designed to accommodate non-uniform traffic distributions.
    This approach enables satellites to allocate more time slots to high-demand areas while providing less frequent service to sparsely populated regions.
    Such adaptability is particularly crucial for LEO satellites, as their rapid orbital motion induces significant variations in traffic demand within their coverage footprint, necessitating dynamic adjustments \cite{Beam-hopping_LEO_oppertunity}. 
    Moreover, beam-hopping offers substantial hardware and cost advantages.
    By illuminating only a subset of cells at any given time, satellites can serve large geographic areas using limited active transceiver chains.
    This efficient hardware utilization facilitates satellite miniaturization, reduces power consumption, and lowers operational costs, which are critical for LEO satellite systems.
    Due to these advantages, beam-hopping has been successfully implemented in commercial satellite systems \cite{Phase_array_for_satellite}.}

    Satellite communication has recently evolved to offer direct-to-cell services for handheld devices and  support for growing IoT connectivity demand.    
    For instance, Starlink has commercialized direct-to-cell services in the United States and New Zealand, with plans to provide IoT services \cite{Starlink_direct-to-cell}.
    However, due to the high satellite altitude and expansive coverage areas, the number of supported devices significantly exceeds that of terrestrial networks, necessitating massive connectivity support \cite{Massive_access_in_space}.
    Additionally, satellites experience long propagation delays due to large distances to the ground (e.g., 2-10 ms one-way propagation delay from LEO satellites).
    Consequently, conventional grant-based transmission becomes inefficient, as the overhead of the handshaking procedure increases due to both massive connectivity and long propagation delays.

    Grant-free random access offers a promising solution to address these challenges \cite{Grant-free_3GPP}.
    This technique eliminates handshaking phases, allowing pilot and data signals to be transmitted simultaneously without requiring dedicated radio resources.
    It reduces access latency and enhances the system's ability to accommodate a large number of devices efficiently, making it well-suited for massive connectivity in satellite networks.
    Motivated by these advantages, prior studies have explored integrating grant-free random access into satellite communications \cite{Grant-free_satellite}.
    However, existing approaches primarily rely on single-beam satellites or fully digital beamforming architectures, which are impractical due to stringent hardware complexity and processing limitations.
    \textcolor{black}{
    Furthermore, no prior work has addressed the unique challenges of integrating beam-hopping with grant-free random access, where the unpredictable nature of device activity shifts optimization objectives from traditional rate-centric metrics to success transmission probability.
    Additionally, the beam-hopping pattern must be designed not only to manage inter-cell interference but also to mitigate the inherent risk of collisions.
    }
    To address this gap, this paper presents a novel beam-hopping pattern design for LEO satellite systems with grant-free random access.


    \subsection{Related Works}
    Beam-hopping techniques have demonstrated potential to enhance system performance in both downlink \cite{Beam-hopping} and uplink \cite{Beam-hopping_uplink} scenarios.
    \textcolor{black}{For example, it has been reported that beam-hopping achieves superior performance compared to multi-color frequency reuse when traffic demand is highly imbalanced, while maintaining competitive performance under uniform traffic conditions \cite{Beam-hoping_versus_multi-color}.}
    Despite this advantage, optimizing beam-hopping patterns remains challenging due to the integer programming formulation.
    To address this, numerous studies have developed algorithms that dynamically adapt to non-uniform traffic demands across cells.
    Early work employed a genetic algorithm \cite{Beam-hopping_genetic}, while reinforcement learning has been applied given its ability to handle binary-constrained optimization \cite{Beam-hopping_RL}.
    Beam-hopping has also been explored in various communication system configurations. 
    For instance, its potential benefits in non-orthogonal multiple access (NOMA) scenarios were analyzed in \cite{Beam-hopping_NOMA_satellite}.
    To address limited coverage of individual LEO satellites, beam patterns of multiple satellites were jointly scheduled to mitigate both inter-beam and inter-satellite interference \cite{Beam-hopping_LEO_multi-satellite}.
    However, no studies have addressed integrating beam-hopping with grant-free random access, despite its potential for massive connectivity in satellite networks.

    To support a large number of remote devices in satellite communication, grant-free random access has garnered significant research attention.
    Several studies have focused on designing device activity detection algorithms that integrate channel estimation and data detection \cite{Grant-free_LEO,Grant-free_MP_LEO,Grant-free_sparse_bayesian_LEO,Grant-free_MIMO-OTFS}.   
    Specifically, in \cite{Grant-free_LEO,Grant-free_MP_LEO}, compressed sensing-based receivers were proposed for joint device detection and channel estimation. In \cite{Grant-free_sparse_bayesian_LEO,Grant-free_MIMO-OTFS}, orthogonal time-frequency-space (OTFS) modulation was considered,  which leverages signal sparsity in the delay-Doppler domain.
    Access control and resource allocation for satellite random access have also been studied to accommodate the unique characteristics of satellite communication.
    In \cite{Random_access_protocol_learning}, a multi-satellite random access protocol was proposed to maximize throughput while minimizing collision rates by employing a multi-agent reinforcement learning model to select serving satellites.
    In \cite{Random_access_resource_allocation}, frequency channel allocation and access probabilities were optimized based on device traffic prediction. 
    \textcolor{black}{However, these prior works exhibit limitations due to reliance on single-beam configurations.
    This approach limits system capacity by preventing spatial frequency reuse and adapting to varying traffic demands, causing resource inefficiency.
    While some works consider multi-beam systems to address these issues, they often depend on fully digital beamforming architectures.
    These approaches, requiring dedicated a radio frequency (RF) chain for each antenna element, face practical limitations in scalability, cost, and power consumption \cite{Holographic_Metasurface_LEO}, making them challenging to implement in large-scale satellite systems.}
    

    \subsection{Our Contributions}
    \textcolor{black}{In this paper, we investigate beam-hopping pattern design for multi-beam LEO satellite systems employing grant-free random access. Specifically, we propose novel beam-hopping pattern design methods that effectively manage inter-cell interference and mitigate collision risks inherent to grant-free random access scenarios where device distribution within the satellite coverage area is imbalanced.} 
    The main contributions of this paper are summarized as follows:

    \begin{itemize}
        \item \textcolor{black}{We formulate a beam-hopping pattern design problem for multi-beam LEO satellite systems operating with grant-free random access.
        Unlike existing beam-hopping studies focusing on data rate optimization, our objective is to maximize the minimum successful transmission probability across serving cells, which is the paramount indicator for systems with sporadic and unpredictable device activity.
        This formulation rigorously accounts for the inherent randomness of grant-free transmissions and the limited satellite beam capacity by characterizing success probability through collision avoidance and decoding success metrics.}

        \item 
        \textcolor{black}{We propose novel beam-hopping pattern design algorithms based on alternating optimization (AO), tailored specifically for multi-beam LEO satellite systems with grant-free random access. A key innovation of our approach is the introduction of an auxiliary variable that enables the decoupled optimization of two critical performance metrics: the collision avoidance rate and the decoding success probability. While the collision avoidance rate can be efficiently maximized using a bisection method, optimizing the decoding success probability is significantly more challenging due to the presence of binary decision variables. To overcome this challenge, we develop alternating direction method of multipliers (ADMM) methods which manage the binary constraint either through simple rounding operations or by applying an $\ell_2$-box constraint.} 

        
        \item Through extensive simulations, we demonstrate the superiority of the proposed beam-hopping pattern design algorithms over existing benchmarks.
        Our results confirm that the proposed algorithms achieve a balance between overall performance and fairness.
        Furthermore, we analyze the impact of varying scales of satellite systems, showing the robustness of the proposed algorithms.

    \end{itemize}

	\begin{figure*}[t]
    	\centering
    	{\epsfig{file=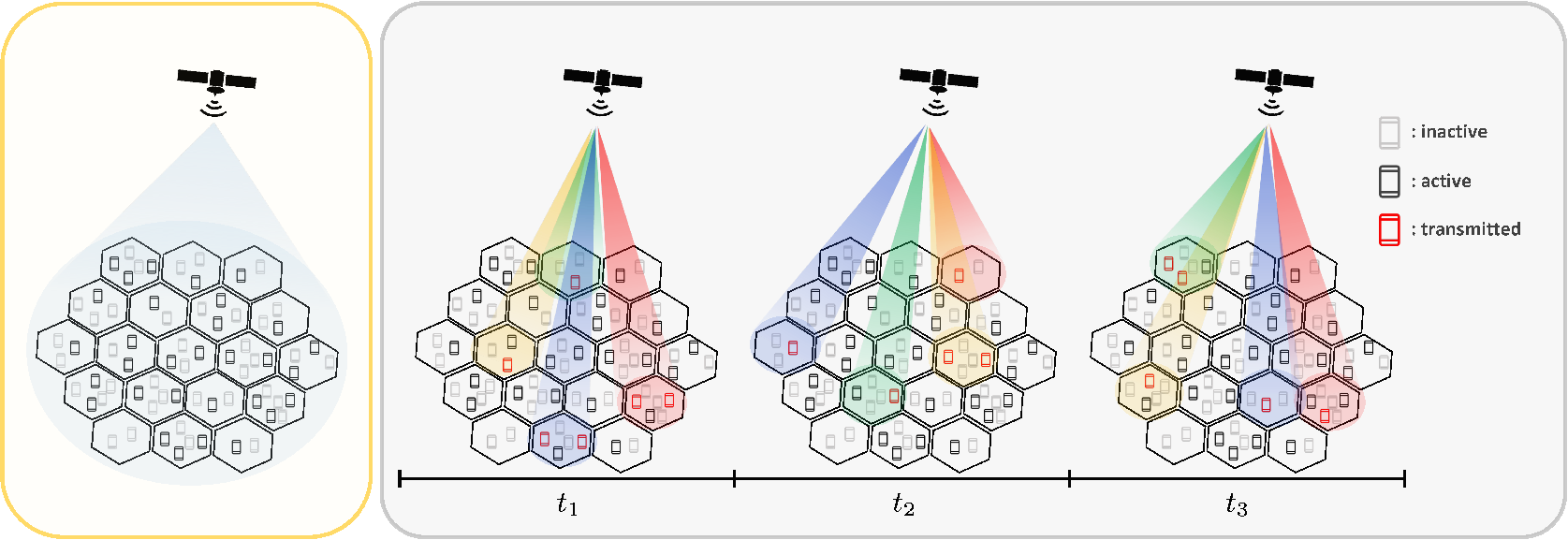, width=16cm}} \vspace{-1mm}
        \caption{Overview of a LEO satellite communication system with grant-free random access, featuring a satellite equipped with beam-hopping capabilities.} \vspace{-3mm}
    	\label{fig:framework}
    \end{figure*}

    \section{System Model}\label{Sec:System Model}

    \subsection{Communication Scenario}
    Consider a LEO satellite communication system with grant-free random access, where the system consists of multi-beam satellites with beam-hopping capabilities. 
    In this scenario, the Earth is gridded into hexagonal cells following the discrete global grid system (DGGS), and the satellite serves a region consisting of $N_c$ cells, indexed by $\mathcal{N}_c \triangleq \{1, 2, \dots, N_c\}$. 
    Each cell $i\in\mathcal{N}_c$ contains $N_i$ single-antenna devices.
    In this work, grant-free random access is adopted, where devices in each cell transmit data directly to the satellite without establishing any connections. 
    Most devices remain inactive, while only those with data to transmit become active.
    The activation probability of each device in cell $i\in\mathcal{N}_c$, denoted by $\alpha_i$, varies across cells and can be inferred by the satellite based on historical records.
    Devices are assumed to have a global navigation satellite system (GNSS); thereby, location reports during registration provide the information of $N_i$ to the satellite.
    
    Multi-beam satellites generate $N_b$ beams simultaneously, illuminating only a subset of the $N_c$ cells (i.e., $N_b < N_c$). 
    To enhance spectrum utilization, each beam employs full frequency reuse, leveraging the entire available bandwidth while causing inter-beam interference in neighboring cells.
    Beam illumination patterns represent the illuminated cells at each time slot and are adjusted dynamically over a beam-hopping window $T_w$, which is divided into $N_{slot}$ time slots of duration $T_{slot}=T_w/N_{slot}$. 
    Each time slot is indexed by $t\in\mathcal{N}_{slot}\triangleq \{1,2,\ldots,N_{slot}\}$.
    The beam illumination pattern for time slot $t$ is denoted as $\mathbf{x}^t$, where $\mathbf{x}^t=[x_1^t, x_2^t,\dots, x_{N_c}^t]^{\sf T}\in \{0,1\}^{N_c}$, and $x_i^t=1$ indicates that cell $i$ is illuminated at time slot $t$.
    Then, the beam illumination patterns across the beam-hopping window are represented by a beam-hopping pattern matrix $\mathbf{X}$ defined as $\mathbf{X} = \left[\mathbf{x}^1, \mathbf{x}^2, \dots, \mathbf{x}^{N_{slot}}\right]$.

    \textcolor{black}{The beam-hopping framework for grant-free random access is illustrated in Fig.~\ref{fig:framework}.}
    Before the beam-hopping window begins, the satellite broadcasts a control signal containing the beam-hopping pattern and synchronization information to all served cells.
    Propagation delay is assumed to be compensated by employing a time-advance technique based on the synchronization information.
    When a device is active, it randomly selects one of the time slots allocated to its cell, where the device is located, and transmits its packet within one of the resource blocks for the selected time slot. 
    Based on its beam-hopping pattern, the satellite illuminates designated cells, receives signals from devices within those cells, checks for collisions, and decodes messages.
    \textcolor{black}{
    For resource modeling, let $N_R$ be the number of available resource blocks per time slot, which are orthogonally divided.
    Active devices choose one of the resource blocks, which become available only if the cell is illuminated (i.e., $x_i^t=1$) within a given time slot of the beam-hopping window.
    Then, the probability that a device in cell $i$ is active and chooses a given resource block from the total available resource blocks within the entire beam-hopping window is given by $\alpha_i/(N_R\sum_{t=1}^{N_{slot}}x_i^t)$. 
    The number of active devices transmitting on a given resource block over the entire beam-hopping window follows a binomial distribution given by}
    \begin{align}\label{eq:Active_devices_Binomial}
        k_i\sim\mathrm{Bino}\left(N_i,\frac{\alpha_i}{N_R\sum_{t=1}^{N_{slot}}x_i^t}\right).
    \end{align}

    Due to the high Doppler shift in LEO satellite communications, the channel between each device and the satellite has a short coherence time, making instantaneous channel state information (iCSI) unavailable in practical systems. 
    Instead, slowly-varying statistical CSI (sCSI) is assumed to be available at the satellite, which includes average received power and angles between the satellite and the devices.
    The average channel gain between the beam illuminating cell $i$ and the device in cell $j$ is given by 
    \begin{align}
       g_{ij} = \frac{G_tG_r(\theta_{ij})}{\left(4\pi {d_{j}}/{\lambda}\right)^2},
    \end{align}
    where $G_t$ is a transmit antenna gain of devices, $G_r(\theta)$ is the satellite's received antenna gain, 
    $d_j$ represents the distance between the satellite and cell $j$, and $\theta_{ij}$ is the angle of arrival (AoA) at the satellite illuminating cell $i$ from a device in cell $j$.
    For simplicity, devices within each cell are assumed to experience the same channel gain.
    

    \subsection{Traffic Demand Model}\label{Sec:Traffic_demand_model}
    \textcolor{black}{The distribution of traffic demands plays a crucial role in designing beam-hopping patterns, as it guides the allocation of beam resources to prioritize high-demand regions and dynamically adapt to diverse traffic conditions across urban, rural, and maritime scenarios. While scenarios with relatively uniform traffic can be adequately modeled using uniform or Poisson distributions, realistically modeling high-contrast demand scenarios remains challenging. Population datasets naturally exhibit strong spatial variation, often covering both densely populated (e.g., urban) and sparsely populated (e.g., rural) areas within a satellite’s coverage footprint. However, publicly available datasets are limited and primarily focused on geostationary satellite systems, making them unsuitable for LEO satellite studies. To address this gap, we construct a dedicated population dataset tailored specifically for LEO satellite systems. Based on this dataset, we propose a population-aware traffic demand model that effectively captures realistic spatial demand variations for LEO satellite systems.} 

    
    To facilitate the cell-based modeling, we use the H3 DGGS developed by Uber, which partitions the Earth's surface into hexagonal cells \cite{H3}.
    H3’s seamless grid structure and hierarchical cell sizes make it particularly suitable for LEO satellite applications. 
    At resolution level 4, H3 divides the Earth into 288,122 cells, each with an average radius of 23.73 km, which represents an appropriate size for LEO satellite beams\footnote{\textcolor{black}{The chosen H3 resolution aligns with 3GPP documents \cite{3GPP-NTN_study} for LEO satellites, which specify a beam diameter of approximately 50 km (25 km radius) and 3dB beamwidth around 4.41$^\circ$.}}.
    Additionally, its open-source availability enhances its utility in this study.

    We consider two primary scenarios for modeling satellite communication demand.
    The first is ubiquitous communication, where satellites provide connectivity in remote areas such as mountains, deserts, oceans, and polar regions. 
    The demand in this scenario follows a uniform distribution, modeled as $u_i\sim \mathcal{U}(0.5, 1.5)$.
    The second is population-centric communication, where communication demand is determined by the population density of each cell. 
    We use GPWv4 population data provided by NASA’s SEDAC \cite{GPWv4_NASA}, which maps estimated 2020 population distributions onto a spherical grid.
    This data is then remapped to H3 cells and yields the estimated population of cell $i$ denoted as $P_i$.
    The demand $p_i$ for the second scenario is set as $p_i = P_i^\beta/P_{avg}$, where $\beta$ is a population scaling factor with $0 \leq \beta \leq 1$, and $P_{\text{avg}}$ is the average population across cells.

    In satellite communications, traffic demand partially follows the population distributions while other portion contribute independently of population.
    Therefore, we propose a hybrid traffic demand model, in which the ubiquitous and population-centric demand models are linearly combined as follows:
    \begin{align}\label{eq:traffic_demand_model}
        N_i=N_{avg}(\eta u_i + (1-\eta){p_i}),
    \end{align}
    where $N_{avg}$ indicates the average traffic demand of cells and $\eta\in[0,1]$ adjusts the weight of the ubiquitous scenario relative to the population-centric scenario.
    When $\eta = 1$, the model accounts solely for the ubiquitous scenario, while when $\eta = 1$, the model becomes purely population-centric. 

                    
                 
    \section{Problem Formulation for Beam-Hopping Pattern Design}\label{Sec:Problem Formulation}
    In this section, we formulate a beam-hopping pattern design problem for the LEO satellite communication system with grant-free random access described in Sec.~\ref{Sec:System Model}. 
    The formulation aims to maximize the minimum success transmission probability across all cells, while considering the satellite's beam generation capacity.


    \subsection{Success Transmission Probability}\label{Sec:Probability}
    In the system described in Sec.~\ref{Sec:System Model}, successful transmission is guaranteed when collisions are avoided and the received signals are decoded without error.
    These two events are statistically independent, implying that the probability of successful transmission can be represented\footnote{In this work, we neglect the effect of automatic repeat request (ARQ) errors for simplicity.} as the product of the collision avoidance rate and decoding success probability. 
    Each factor can be characterized as described below.

    \subsubsection{Collision avoidance rate} 
    In the grant-free random access scenario, each active device randomly selects one of the assigned time slots and resource blocks, as described in Sec.~\ref{Sec:System Model}.
    Collisions occur when multiple devices in the same cell select the same resource block in the same time slot for access.
    \textcolor{black}{Therefore, following the derivation presented in \cite{Resource_hopping_grant-free}, the collision avoidance rate at cell $i\in\mathcal{N}_c$, $P_{a,i} \in [0,1]$, is given by}
    \begin{align}
        P_{a,i}(\mathbf{x}_i) = \left(1-\frac{\alpha_i}{N_R\sum_{t=1}^{N_{slot}}x_i^t}\right)^{N_i-1},
    \end{align}
    \textcolor{black}{where $\mathbf{x}_i = [x_i^1, x_i^2,\cdots, x_i^{N_{slot}}]^{\sf T}$ represents the beam-hopping pattern for cell $i$.}

    \subsubsection{Decoding success probability}
    Even when devices access satellites without collisions, packet decoding may fail due to inter-beam interference and high pathloss.
    These are captured by assuming that the packets are successfully decoded only when the signal-to-noise-plus-interference ratio (SINR), including the inter-beam interference effects, exceeds a threshold $\gamma_{th} > 0$.
    Then, the decoding success probability is expressed as
    \begin{align}\label{eq:Decoding_success_probability}
        P_{d,i}(\mathbf{X}) = \frac{1}{\sum_{t'=1}^{N_{slot}}x_i^{t'}}\sum_{t=1}^{N_{slot}}x_i^t \mathbb{P}\left[\frac{g_{ii}}{\sum_{j\neq i}x_j^t k_j g_{ij}+ \rho^{-1}} >\gamma_{th}\right]\!,
    \end{align}
    where \textcolor{black}{$\mathbb{P}[\cdot]$ denotes the probability function and
    $\rho$ is the signal-to-noise ratio (SNR) of the system.}
    \textcolor{black}{
    The inter-beam interference $\sum_{j\neq i}x_j^t k_j g_{ij}$ sums interference from all other active cells, determined by the beam indicators $\{x_j^t\}$, the random numbers of active devices $\{k_j\}$, and the average channel gains $\{g_{ij}\}$.
    The beam indicators $\{x_j^t\}$ are intertwined with this inter-beam interference, as the interference at one cell depends on the beam indicators $\{x_j^t\}$ for all other cells.}

    \textcolor{black}{Unlike inter-beam interference, intra-cell interference, which occurs when multiple devices within the same cell transmit on the same resource, is not accounted for the decoding success probability but is instead captured solely by the collision avoidance rate $P_{a,i}$. This is because devices in the same cell are typically at similar distances from the high-altitude satellite, resulting in received signals with comparable power levels. Consequently, the receiver is unable to distinguish and decode any individual signal, leading to collisions that prevent successful transmission for all involved devices. As a result, only collision-free transmissions are considered for decoding, and their SINRs are primarily affected by inter-beam interference.}

    \subsection{Problem Formulation}\label{Sec:Problem}
    Analysis in Sec.~\ref{Sec:Probability} reveals that the success transmission probability for cell $i$, given beam-hopping pattern matrix $\mathbf{X}$, is characterized as $P_{suc,i}(\mathbf{X})= P_{a,i}(\mathbf{x}_i)P_{d,i}(\mathbf{X})$, which is the product of collision avoidance rate $P_{a,i}(\mathbf{x}_i)$ and decoding success probability $P_{d,i}(\mathbf{X})$.
    Using this result, we formulate a beam-hopping pattern design problem to determine the beam-hopping pattern matrix that maximizes the minimum success transmission probability across all cells, considering the satellite’s beam generation capacity. 
    The optimization problem is given below. 
    \begin{subequations}
    \begin{align}\label{eq:P_0}
    \mathcal{P}_{0}:\underset{\mathbf{X}}{\mathrm {max}} ~ &\min_{i\in\mathcal{N}_c} P_{suc,i}(\mathbf{X})
    \\
    ~\text {s.t.}~&{P_{suc,i}(\mathbf{X})}= P_{a,i}(\mathbf{x}_i)P_{d,i}(\mathbf{X}),~~ \forall i\in \mathcal{N}_c,
    \label{eq:P_0_C1}\\
    &\sum_{i\in\mathcal{N}_c} x_{i}^t \leq N_b, ~~  \forall t\in\mathcal{N}_{slot}, 
    \label{eq:P_0_C2}\\
    &x_{i}^t \in \{0,1\}, ~~ \forall i\in\mathcal{N}_c, ~ \forall t\in\mathcal{N}_{slot},
    \label{eq:P_0_C3}\\
    &\sum_{t=1}^{N_{slot}} x_{i}^t \geq 1, ~~ \forall i\in\mathcal{N}_c. \label{eq:P_0_C4}
    \end{align}
    \end{subequations}
    In the above problem, 
    the constraint in \eqref{eq:P_0_C2} limits the available number of beams because the satellite can illuminate at most $N_b$ cells simultaneously.
    The constraint in \eqref{eq:P_0_C3} restricts beam illumination indicators $x_i^t$ to binary values, and the constraint in \eqref{eq:P_0_C4} ensures that each cell $i\in\mathcal{N}_c$ is illuminated at least once within the beam-hopping window.


    \section{Proposed Beam-Hopping Pattern Design Algorithm}\label{Sec:Proposed}
    In this section, we propose a beam-hopping pattern design algorithm that employs an AO approach to solve the optimization problem $\mathcal{P}_{0}$ formulated in Sec.~\ref{Sec:Problem}.

    \subsection{Reformulation of Optimization Problem}
    Directly solving the problem $\mathcal{P}_{0}$ is challenging, as it is non-convex and NP-hard due to intractable expressions in the decoding success probability $P_{d, i}(\mathbf{X})$ and the binary constraint.
    \textcolor{black}{
    The intractability of $P_{d,i}(\mathbf{X})$ in \eqref{eq:Decoding_success_probability} arises from its inter-beam interference term, which is a weighted sum of independent binomial random variables $\{k_j\}$. 
    This makes the objective function stochastic and lacks simple closed-form expressions.
    To make the problem tractable, we introduce a lower bound on the decoding success probability via the Markov inequality.
    Specifically, the SINR condition from \eqref{eq:Decoding_success_probability} is rewritten as an inequality on the non-negative random interference term $Z_t = \sum_{j\ne i}x_{j}^{t}k_{j}g_{ij}$.
    We then apply the Markov inequality in the form $\mathbb{P}(Z_t < a) \ge 1 - \frac{\mathbb{E}[Z_t]}{a}$, where the expectation $\mathbb{E}[Z_t]$ is calculated using properties of the binomial random variable $k_j$ defined in \eqref{eq:Active_devices_Binomial}.
    This approximation simplifies the optimization by replacing intractable probabilistic terms with tractable analytical expressions.
    Based on the above strategy, we derive a lower bound of $P_{d,i}(\mathbf{X})$ as follows:
    }
    \begin{align}\label{eq:LB_P_di}
        &P_{d, i}(\mathbf{X})   \nonumber \\
        &\geq  \frac{1}{\sum_{t^\prime=1}^{N_{slot}}x_i^{t^\prime}}\sum_{t=1}^{N_{slot}}x_i^t \left(1-\sum_{j\neq i} \frac{x_j^t g_{ij} N_j \alpha_j}{\sum_{t^\prime=1}^{N_{slot}}x_j^{t^\prime} N_R(\frac{g_{ii}}{\gamma_{th}}-\rho^{-1})}\right) \nonumber \\
        &=  1-  \sum_{t=1}^{N_{slot}}\sum_{j\neq i}  \frac{ x_i^t   x_j^t g_{ij} N_j \alpha_j}{b_i b_j  N_R(\frac{g_{ii}}{\gamma_{th}}- \rho^{-1} )}   \triangleq P_{d,i}^{low}(\mathbf{X}),
    \end{align}
    where $b_i=\sum_{t=1}^{N_{slot}}x_i^t$ denotes the total number of beam illuminations allocated to cell $i$ and $\rho$ is the SNR of the system.
    Introducing a new notation $b_i$ allows rewriting the collision avoidance rate $P_{a,i}({\bf x}_i)$ as a function of $b_i$ as follows:
    \begin{align}\label{eq:P_ai2}
        P_{a,i}(b_i) = \left(1-\frac{\alpha_i}{N_Rb_i}\right)^{N_i-1}.
    \end{align}
    From \eqref{eq:LB_P_di} and \eqref{eq:P_ai2}, we determine the lower bound of the success transmission probability for cell $i$, given by 
    \begin{align}
    P_{suc,i}(\mathbf{X}) \geq P_{suc,i}^{low}(\mathbf{X},{\bf b}) = P_{a,i}(b_i)P_{d,i}^{low}(\mathbf{X}),
    \end{align}
    where ${\mathbf{b}} = [{b}_1,\cdots, {b}_{N_c}]^{\sf T}$.
    Using this lower bound, we reformulate problem $\mathcal{P}_0$ which now aims at maximizing the minimum lower bound of the success transmission probability across all cells:
    \begin{subequations}\label{eq:P_1_formulation}
    \begin{align}
    \mathcal{P}_{1}:\underset{\mathbf{X}, \mathbf{b}}{\mathrm {max}} ~ &\min_{i\in\mathcal{N}_c} P_{suc,i}^{low}(\mathbf{X},{\bf b})
    \\
    \text {s.t.}~&{P_{suc,i}^{low}(\mathbf{X},{\bf b})}= P_{a,i}(b_i)P_{d,i}^{low}(\mathbf{X}),~~ \forall i\in \mathcal{N}_c,
    \\
    &b_{i} = \sum_{t=1}^{N_{slot}}x_i^t, ~~ \forall i\in\mathcal{N}_c,\label{eq:P_1_C2}
    \\
    &b_{i} \geq 1, ~~ \forall i\in\mathcal{N}_c,\\
    &\eqref{eq:P_0_C3},\eqref{eq:P_0_C4}.
    \end{align}
    \end{subequations}
    The reformulation in \eqref{eq:P_1_formulation} clearly shows that variables $\mathbf{X}$ and $\mathbf{b}$ are intertwined.
    Motivated by this, we employ an AO approach to solve the problem $\mathcal{P}_{1}$, which alternately determines $\mathbf{X}$ and $\mathbf{b}$ to maximize the minimum lower bound of the success transmission probability. 

    \subsection{Collision Avoidance Rate Maximization: Bisection Method} 
    \textcolor{black}{In our AO approach, we relax the hard constraint in \eqref{eq:P_1_C2} and treat $b_i$ as free variables to be optimized, while using the fixed ${\bf X}$ from the previous iteration to calculate the term $P_{d,i}^{low}({\bf X})$ and initialize $b_i$.
    Based on this strategy, we formulate a sub-problem to maximize a minimum \textit{weighted} collision avoidance rate for the fixed $\mathbf{X}$ as follows:}
    \begin{subequations}\label{eq:P_2}
    \begin{align}
    \mathcal{P}_{2}:\underset{\mathbf{b}}{\mathrm {max}} &~ \min_{i\in\mathcal{N}_c} P_{a,i}(b_i)P_{d,i}^{low}(\mathbf{X})
    \\
    \text {s.t.}~&\sum_{i\in\mathcal{N}_c} b_i \leq N_{slot}N_b, \label{eq:P_2_C2}
    \\
    &b_i\leq N_{slot}, ~~ \forall i\in\mathcal{N}_c, \label{eq:P_2_C3}
    \\
    & b_i\in\mathbb{Z}_+, ~~ \forall i\in\mathcal{N}_c,\label{eq:P_2_C4}
    \end{align}
    \end{subequations}
    where $P_{d,i}^{low}(\mathbf{X})$ is treated as a constant computed from the given $\mathbf{X}$. 
    To transform the problem $\mathcal{P}_{2}$ into an epigraph form, we introduce the auxiliary variable $\xi\triangleq \min_{i\in\mathcal{N}_c} P_{a,i}(b_i)P_{d,i}^{low}(\mathbf{X})$.
    Then, an equivalent problem is given by 
    \begin{subequations}\label{eq:P_2P}
    \begin{align}
    \mathcal{P}_{2}^\prime:\underset{\mathbf{b}, \xi}{\mathrm {max}} ~ &\xi
    \\
    ~\text {s.t.}~&\left(1-\frac{\alpha_i}{N_Rb_i}\right)^{N_i-1}P_{d,i}^{low}(\mathbf{X}) \geq \xi,~~ \forall i\in \mathcal{N}_c,
   \label{eq:P_2P_C1} \\
    &\eqref{eq:P_2_C2}, \eqref{eq:P_2_C3},\eqref{eq:P_2_C4}.
    \end{align}
    \end{subequations}
    The constraint \eqref{eq:P_2P_C1} can be rewritten as
    \begin{align}
         b_i\geq \frac{\alpha_i}{\left(1-\exp{\frac{\log\xi-\log P_{d,i}^{low}(\mathbf{X})}{N_i-1}}\right)N_R}\triangleq \bar{f}_i(\xi),~\forall i\in\mathcal{N}_{c}.
    \end{align}
    For a given of $\xi$, the variable $\mathbf{b}$ satisfying the constraints \eqref{eq:P_2P_C1}, \eqref{eq:P_2_C3}, and \eqref{eq:P_2_C4} is determined as
    \begin{align} 
        b_i = \min(\max(\lceil \bar{f}_i(\xi) \rceil, 1), N_{slot}) \triangleq f_i(\xi) \label{eq:Bisection b}.
    \end{align}
    If the the variable $\mathbf{b}$ in \eqref{eq:Bisection b} satisfies the constraint \eqref{eq:P_2_C2}, then the given $\xi$ is valid and can be set as the lower bound of the optimal $\xi^*$.
    Conversely, if infeasible, $\xi$ becomes the upper bound of $\xi^*$.
    Since we can verify feasibility and calculate $\mathbf{b}$ for a given $\xi$ when feasible, the problem $\mathcal{P}_2^\prime$ can be solved using a bisection method, which iteratively updates the upper and lower bounds of $\xi$.
    The bisection method for maximizing the collision avoidance rate is summarized in Algorithm~\ref{alg:Bisection}.

    We characterize the uniqueness and properties of the optimal point of the problem $\mathcal{P}_2^\prime$, as given in the following lemma:
    \begin{lem}\label{lem:bisection uniqueness}
        If $\mathcal{P}^\prime_2$ is feasible and $\lim_{\xi\rightarrow a+}\sum_i f_i(\xi)-\sum_i f_i(a) \leq 1$, then the optimal point $(\mathbf{b}^*, \xi^*)$ is unique and satisfies $\sum_{i\in\mathcal{N}_c}b_i^*= N_{slot}N_b$.
    \end{lem}
    \begin{IEEEproof} 
       See Appendix \ref{apdx:bisection uniqueness}. 
    \end{IEEEproof}
    \noindent Note that the condition $\lim_{\xi\rightarrow a+}\sum_i f_i(\xi)-\sum_i f_i(a) \leq 1$ in Lemma \ref{lem:bisection uniqueness} holds in practice due to the real-valued nature of $P_{d,i}^{low}(\mathbf{X})$. Based on this result, we analyze the convergence of the bisection method in Algorithm~\ref {alg:Bisection}. The result is given in the following theorem:
    \begin{thm}\label{lem:bisection convergence}
        Algorithm \ref{alg:Bisection} converges to the optimal solution $\xi^*$ of problem $\mathcal{P}_2^\prime$.
    \end{thm}
    \begin{IEEEproof} 
       See Appendix \ref{apdx:bisection convergence}.
    \end{IEEEproof}
    
    \begin{algorithm}[t]
        \caption{Bisection Method}\label{alg:Bisection}
        {\small \begin{algorithmic}[1]
            \REQUIRE $P_{d,i}^{low}(\mathbf{X}), \alpha_i, N_i, N_{slot}, \text{ and } N_b$ 
            \STATE  \textbf{Initialize:} $\xi_{\ell} \gets 0, \xi_{u}\gets P_{d,i}^{low}(\mathbf{X})$.
            \REPEAT
            \STATE $\xi_{\mathrm{mean}}\gets\frac{\xi_{\ell}+\xi_{u}}{2}$.
            \STATE Update $b_i$ by \eqref{eq:Bisection b}$, ~ \forall i\in\mathcal{N}_c$.
            \IF {$\sum_{i\in\mathcal{N}_c}b_i \leq N_{slot}N_b$}
            \STATE $\xi_{\ell} \gets \xi_{\mathrm{mean}}$.
            \ELSE
            \STATE $\xi_{u} \gets \xi_{\mathrm{mean}}$.
            \ENDIF
            \UNTIL
            \STATE For given $\xi_{\ell}$, update $b_i$ by \eqref{eq:Bisection b} $, ~ \forall i\in\mathcal{N}_c$.
            \ENSURE $\mathbf{b}$
        \end{algorithmic}}
    \end{algorithm}
    
    \subsection{Decoding Success Probability Maximization: ADMM}
    We now describe the strategy for optimizing $\mathbf{X}$ while keeping $\mathbf{b}$ fixed.
    Since the collision avoidance rate is solely determined by $\mathbf{b}$, the rate $P_{a,i}(b_i)$ can be treated as a constant computed from $\mathbf{b}$. 
    This allows focusing only on optimizing the decoding success probability.
    \textcolor{black}{Unfortunately, maximizing the decoding success probability is still challenging due to the intractability of the min operation.
    To circumvent this challenge, we transform the objective into a summation of the success transmission probabilities across cells as follows:}
    \begin{subequations}
    \begin{align}
        \mathcal{P}_{3}:\underset{\mathbf{X}}{\mathrm {max}} ~ &\sum_{i\in\mathcal{N}_c}P_{a,i}(b_i)P_{d,i}^{low}(\mathbf{X})
        \\
        \text {s.t.}~& \sum_{t=1}^{N_{slot}}x_i^t=b_i, ~~\forall i \in \mathcal{N}_c, \label{eq:P_3_C1}
        \\
        &\eqref{eq:P_0_C2},\eqref{eq:P_0_C3}.
    \end{align}
    \end{subequations}
    \textcolor{black}{While this sub-problem maximizes the summation, fairness in beam illumination across cells is preserved by the constraint in \eqref{eq:P_3_C1} because the beam allocation vector $\mathbf{b}$ in this constraint has already been optimized to maximize the minimum success transmission probability by solving the sub-problem $\mathcal{P}_{2}$. The effectiveness of this approach will be further validated through simulation results in Sec.~V.} The objective in the sub-problem $\mathcal{P}_{3}$ can be rewritten in quadratic form:
    \begin{align}
        \!\!\sum_{i\in\mathcal{N}_c}P_{a,i}(b_i)P_{d,i}^{low}(\mathbf{X})  {=} \sum_{i\in\mathcal{N}_c}P_{a,i}(b_i){-}\sum_{t=1}^{N_{slot}}(\mathbf{x}^t)^{\sf T}\tilde{\mathbf{G}}\mathbf{x}^t,
    \end{align}
    where 
    \begin{align}
        \{\tilde{\mathbf{G}}\}_{i,j} = 
        \begin{cases}
        \frac{P_{a,i}(b_i)g_{ij} N_j \alpha_j}{b_ib_jN_R(\frac{g_{ii}}{\gamma_{th}}-\rho^{-1})}, &\quad i\neq j\\
        0, &\quad i=j.
        \end{cases} \label{eq: quadractic matrix}
    \end{align}
    To ensure the convexity of the objective, we make the matrix $\tilde{\mathbf{G}}$ positive semidefinite by applying 
    \begin{align*}
        \bar{\mathbf{G}} = \frac{1}{2}(\tilde{\mathbf{G}}+\tilde{\mathbf{G}}^T),~~\text{and}~~
        \mathbf{G} = \bar{\mathbf{G}}-\lambda_{\min}(\bar{\mathbf{G}})\mathbf{I}, 
    \end{align*}
    where $\lambda_{\min}(\cdot)$ denotes the smallest eigenvalue of a matrix. 
    Note that the matrix $\mathbf{G}$ satisfies $\sum_{t=1}^{N_{slot}}(\mathbf{x}^t)^{\sf T}\mathbf{G}\mathbf{x}^t=\sum_{t=1}^{N_{slot}}(\mathbf{x}^t)^{\sf T}\tilde{\mathbf{G}}\mathbf{x}^t-\lambda_{\min}(\bar{\mathbf{G}})\mathbf{I}$ from the constraint in \eqref{eq:P_3_C1}.
    Then, the sub-problem $\mathcal{P}_3$ is equivalently reformulated as
    \begin{subequations}
    \begin{align}
        \mathcal{P}_{3}^{\prime}:\underset{\mathbf{X}}{\mathrm {min}} ~ &\sum_{t=1}^{N_{slot}}(\mathbf{x}^t)^{\sf T}\mathbf{G}\mathbf{x}^t
        \\
        ~\text {s.t.}~& \eqref{eq:P_0_C2},\eqref{eq:P_0_C3}, \eqref{eq:P_3_C1},
    \end{align}
    \end{subequations}
    which is a constrained binary quadratic optimization problem. 

    Although the unconstrained binary quadratic optimization problem has been extensively studied \cite{UBQP}, the constrained binary quadratic optimization problem remains an open challenge.
    \textcolor{black}{Well-known approaches, such as branch-and-bound \cite{branch_and_bound} and semidefinite relaxation \cite{Semidefinite_programming_relaxation}, are available;  however, their application is impractical in our case due to the large dimensionality of the optimization variable $\mathbf{X}$.}
    Instead, we employ the ADMM algorithm, which combines the augmented Lagrangian method and dual ascent to solve the problem $\mathcal{P}_{3}^{\prime}$.
    While ADMM is widely used for convex optimization, its extension to non-convex problems has gained considerable attention, with recent findings highlighting its advantages \cite{ADMM_nonconvex,ADMM_binary}.
    To employ the ADMM algorithm, we define two constraint sets $\Omega_1$ and $\Omega_2$ as 
    \begin{align}
        \Omega_1&=\{\mathbf{X}:\mathbf{X}\in\{0,1\}^{N_{c}\times N_{slot}}\}, \\
        \Omega_2&= \{\mathbf{X}:\mathbf{X}^{\sf T}\mathbf{1}=N_{b}\cdot \mathbf{1}, \mathbf{X}\mathbf{1}=\mathbf{b}\},
    \end{align}
    which represent the binary and satellite beam constraints, respectively.
    We then derive the ADMM update step based on the augmented Lagrangian given by
    \begin{align} \label{eq:Augmented Lagrangian}
        &\mathcal{L}(\mathbf{X},\mathbf{Z}_1, \mathbf{Z}_2, \mathbf{Y}_1, \mathbf{Y}_2) \nonumber\\
        &=
        \sum_{j=1}^{N_{slot}}\mathbf{x}_j^{\sf T}\mathbf{G}\mathbf{x}_j + I_{\Omega_1}(\mathbf{Z}_1)+I_{\Omega_2}(\mathbf{Z}_2)+\mathrm{Tr}(\mathbf{Y}_1^{\sf T}(\mathbf{X-Z}_1)) \nonumber\\
        &~~~ +\mathrm{Tr}(\mathbf{Y}_2^{\sf T}(\mathbf{X-Z}_2))+\frac{\rho_1}{2}\Vert\mathbf{X-Z}_1\Vert^2_F+\frac{\rho_2}{2}\Vert\mathbf{X-Z}_2\Vert^2_F,
    \end{align}
    where $I_{\Omega}(\cdot)$ denotes the indicator function and $\mathrm{Tr}(\cdot)$ is the matrix trace.
    $\mathbf{Y}_1$ and $\mathbf{Y}_2$ denote dual variables, and parameters $\rho_1$ and $\rho_2$ control the weights of penalty terms.
    The variables $(\mathbf{X}, \mathbf{Z}_1, \mathbf{Z}_2)$ are updated to minimize the augmented Lagrangian as follows:
    \begin{align}
        \mathbf{Z}_1^{k+1} &=\underset{\mathbf{Z}_1\in\Omega_1}{\mathrm{argmin}} ~ \mathrm{Tr}(\mathbf{Y}_1^T(\mathbf{X-Z_1}))+\frac{\rho_1}{2}\Vert \mathbf{X}-\mathbf{Z}_1\Vert^2_F\nonumber\\
        &= \mathrm{round}\left(\mathbf{X}^k+\frac{1}{\rho_1}\mathbf{Y}_1^k \right), \label{eq:round} \\ 
        \mathbf{Z}_2^{k+1} &=\underset{\mathbf{Z}_2\in\Omega_2}{\mathrm{argmin}} ~ \mathrm{Tr}(\mathbf{Y}_2^T(\mathbf{X-Z_2}))+\frac{\rho_2}{2}\Vert \mathbf{X}-\mathbf{Z}_2\Vert^2_F\nonumber\\
        &= \mathbf{X}^k+\frac{1}{\rho_2}\mathbf{Y}_2^k+\frac{\mathbf{1}\boldsymbol{\lambda}^{\sf T}}{\rho_2}+\frac{\boldsymbol{\nu}\mathbf{1}^{\sf T}}{\rho_2} \label{eq:ADMM Z_2},\\
        \mathbf{X}^{k+1} &= (2\mathbf{G}+(\rho_1+\rho_2)\mathbf{I})^{-1} \nonumber \\
        &~~~~\times (\rho_1\mathbf{Z}_1^{k+1}+\rho_2\mathbf{Z}_2^{k+1}-\mathbf{Y}_1-\mathbf{Y}_2), \label{eq:ADMM X}
    \end{align}
    where KKT conditions provide closed-form expressions for updating $\mathbf{Z}_2$ and $\mathbf{X}$, with the associated Lagrange multipliers $\boldsymbol{\nu}$ and $\boldsymbol{\lambda}$ having analytical solutions that can be readily derived.
    \textcolor{black}{
    The update for ${\bf Z}_1$ in \eqref{eq:round} manages the binary constraint by projecting relaxed, continuous variables onto the set $\{0,1\}^{N_c\times N_{slot}}$ through simple rounding operations.
     }
    Following this, the dual ascent is applied to the dual problem, updating $\mathbf{Y}_1$ and $\mathbf{Y}_2$.
    The ADMM steps are summarized in Algorithm \ref{alg:ADMM}.
   
    \begin{algorithm}[t]
        \caption{ADMM}\label{alg:ADMM}
        {\small \begin{algorithmic}[1]
            \REQUIRE $\mathbf{b}, \mathbf{G}, \rho_1, \rho_2, \text{ and }\gamma$
            \STATE  \textbf{Initialize:} $\xi_{\min} \gets 0, \xi_{\max}\gets 1$.
            \REPEAT
            \STATE \!\! $\mathbf{Z}_1^{k+1} \gets \mathrm{round}(\mathbf{X}^k+\frac{1}{\rho_1}\mathbf{Y}_1^k)$.
            \STATE \!\! $\mathbf{Z}_2^{k+1} \gets \mathbf{X}^k+\frac{1}{\rho_2}\mathbf{Y}_2^k+\frac{\mathbf{1}\boldsymbol{\lambda}^{\sf T}}{\rho_2}+\frac{\boldsymbol{\nu}\mathbf{1}^{\sf T}}{\rho_2}$.
            \STATE \!\! $\mathbf{X}^{k+1} \gets (2\mathbf{G}+(\rho_1+\rho_2)\mathbf{I})^{-1}(\rho_1\mathbf{Z}_1^{k+1}+\rho_2\mathbf{Z}_2^{k+1}-\mathbf{Y}_1-\mathbf{Y}_2)$.
            \STATE \!\! $\mathbf{Y}_1^{k+1}\gets \mathbf{Y}_1^k+\gamma\rho_1(\mathbf{X}^{k+1}-\mathbf{Z}_1^{k+1})$.
            \STATE \!\! $\mathbf{Y}_2^{k+1}\gets \mathbf{Y}_2^k+\gamma\rho_2(\mathbf{X}^{k+1}-\mathbf{Z}_2^{k+1})$.
            \UNTIL
            \STATE $\mathbf{X}_{out}\gets \mathrm{round}(\mathbf{X})$.
            \ENSURE $\mathbf{X}_{out}$
        \end{algorithmic}}
    \end{algorithm}

    \subsection{Decoding Success Probability Maximization: $\ell_2$-Box ADMM}\label{Sec:l2-ADMM}
    In the previous subsection, ADMM is employed to find the beam-hopping pattern $\mathbf{X}$ which maximizes the decoding success probability.
    However, the \textit{hard} rounding operations in \eqref{eq:round} make the ADMM updates unstable due to their discrete nature.
    To address this limitation, we replace the binary constraint \eqref{eq:P_0_C3} with an equivalent constraint, which is the intersection between a box and an $\ell_2$-sphere \cite{lp_box_admm}:
    \begin{align} \label{eq:l2_box_intersection}
        \mathbf{x}\in\{0,1\}^n \Leftrightarrow \mathbf{x}\in[0,1]^n \cap \left\{ \mathbf{x}:\left\Vert \mathbf{x}-\frac{1}{2}\mathbf{1} \right\Vert^2_2= \frac{n}{4} \right\}.
    \end{align}
    where $n=N_c\times N_{slot}$ represents the total number of binary variables.
    These continuous constraints induce soft updates instead of hard decision updates in Algorithm \ref{alg:ADMM}.
    Using the equivalence in \eqref{eq:l2_box_intersection}, we reformulate the sub-problem $\mathcal{P}_3$ as follows:
    \begin{subequations}
    \begin{align}
        \mathcal{P}_{4}:\underset{\mathbf{X}}{\mathrm {min}}& ~ \sum_{t=1}^{N_{slot}}(\mathbf{x}^t)^{\sf T}\mathbf{G}\mathbf{x}^t
        \\
        \text {s.t.}~&\eqref{eq:P_0_C2},\eqref{eq:P_0_C3}, \eqref{eq:P_3_C1}, \eqref{eq:l2_box_intersection}.
    \end{align}
    \end{subequations}
    \textcolor{black}{
    This reformulation is key to enhancing the algorithm by eliminating hard rounding operations.
    Instead, it uses projections onto continuous sets, yielding \textit{soft} updates that improve convergence and performance.
    }
    To solve the problem $\mathcal{P}_4$ using the ADMM algorithm, we redefine two constraint sets $\Omega_1$ and $\Omega_2$ as 
    \begin{align}
        \Omega_1&=\{\mathbf{X}:\mathbf{X}\in[0,1]^{N_{c}\times N_{slot}}\}, \\
        \Omega_2&=\left\{\mathbf{X}:\left\Vert \mathbf{X}-\frac{1}{2}\mathbf{1}\mathbf{1}^{\sf T} \right\Vert^2_2 = \frac{N_cN_{slot}}{4}\right\}, 
    \end{align}
    which represent the box and the $\ell_2$-sphere, respectively.
    We then derive the ADMM updates where the augmented Lagrangian is given by
    \begin{align} \label{eq:l2ADMM_augmented_Lagrangian}
        &\!\!\!\!\mathcal{L}(\mathbf{X},\mathbf{Z}_1, \mathbf{Z}_2, \mathbf{Y}_1, \mathbf{Y}_2,\mathbf{y}_3,\mathbf{y}_4)\nonumber\\
        =&\sum_{j=1}^{N_{slot}}\mathbf{x}_j^{\sf T}\mathbf{G}\mathbf{x}_j + I_{\Omega_1}(\mathbf{Z}_1)+I_{\Omega_2}(\mathbf{Z}_2)+\mathrm{Tr}(\mathbf{Y}_1^{\sf T}(\mathbf{X-Z}_1))\nonumber
        \\
        &+\mathrm{Tr}(\mathbf{Y}_2^{\sf T}(\mathbf{X-Z}_2))+\mathbf{y}_3^{\sf T}(\mathbf{X}^{\sf T}\mathbf{1}-N_{b}\mathbf{1})+\mathbf{y}_4^{\sf T}(\mathbf{X1-b})\nonumber
        \\
        &+\frac{\rho_1}{2}\Vert\mathbf{X-Z}_1\Vert^2_F+\frac{\rho_2}{2}\Vert\mathbf{X-Z}_2\Vert^2_F+\frac{\rho_3}{2}\Vert\mathbf{X}^{\sf T}\mathbf{1}-N_{b}\mathbf{1}\Vert^2_2\nonumber
        \\ 
        &+\frac{\rho_3}{2}\Vert\mathbf{X}\mathbf{1}-\mathbf{b}\Vert^2_2,
    \end{align}
    where the dual variables are denoted by $\mathbf{Y}_1, \mathbf{Y}_2, \mathbf{y}_3, \mathbf{y}_4$, and the parameters $\rho_1, \rho_2, \rho_3$ control the weights of penalty terms.
    For simplicity, the same penalty parameter is set for $\mathbf{y}_3$ and $\mathbf{y}_4$. 
    Note that only two auxiliary variables $\mathbf{Z}_1$ and $\mathbf{Z}_2$ are used for the constraints, while the satellite beam constraints are applied to the variable $\mathbf{X}$ rather than using additional auxiliary variables.
    This design choice comes from the observation that a large number of auxiliary variables degrade ADMM performance.
    The projection of $\mathbf{Z}_1$ and $\mathbf{Z}_2$ onto closed sets is performed as follows:
    \begin{align}
        \mathbf{Z}_1^{k+1}&=\underset{\mathbf{Z}_1\in\Omega_1}{\mathrm{argmin}} ~ \mathrm{Tr}(\mathbf{Y}_1^{\sf T}(\mathbf{X-Z_1}))+\frac{\rho_1}{2}\Vert \mathbf{X}-\mathbf{Z}_1\Vert^2_F\nonumber\\
        &=\min(\mathbf{1}_{N_c\times N_{slot}}, \max(\mathbf{X}^k+\frac{\rho_1}{2}\mathbf{Y}^k_1, \mathbf{0}_{N_c\times N_{slot}}),\label{eq:l2ADMM_Z1}\\
        \mathbf{Z}_2^{k+1}&=\underset{\mathbf{Z}_2\in\Omega_2}{\mathrm{argmin}} ~ \mathrm{Tr}(\mathbf{Y}_1^{\sf T}(\mathbf{X-Z_2}))+\frac{\rho_2}{2}\Vert \mathbf{X}-\mathbf{Z}_2\Vert^2_F\nonumber\\
        &=P_{\Omega_2}\left(\mathbf{X}^k+\frac{1}{\rho_2}\mathbf{Y}_2^k\right),\label{eq:l2ADMM_Z2}
    \end{align}
    where 
    \begin{align}
        P_{\Omega_2}(\mathbf{A})=\frac{N_cN_{slot}}{2}\frac{\mathbf{A}-\frac{1}{2}\mathbf{1}_{N_c\times N_{slot}}}{\Vert \mathbf{A}-\frac{1}{2}\mathbf{1}_{N_c\times N_{slot}}\Vert_F}+ \frac{1}{2}\mathbf{1}_{N_c\times N_{slot}}.
    \end{align}
    Next, the variable $\mathbf{X}$ is updated by taking the derivative of \eqref{eq:l2ADMM_augmented_Lagrangian} with respect to ${\bf X}$ and solving for the value that sets it to zero:
    \begin{align} \label{eq:sylvester_eqaution}
        &(2\mathbf{G}+(\rho_1+\rho_2)\mathbf{I}+2\rho_3\mathbf{11}^{\sf T})\mathbf{X}+\mathbf{X}(2\rho_3\mathbf{11}^{\sf T})\nonumber\\
        &=(\rho_1\mathbf{Z}_1+\rho_2\mathbf{Z}_2-\mathbf{Y}_1-\mathbf{Y}_2)\nonumber\\
        &~~~+2(\rho_3N_{b}\mathbf{11}^{\sf T}+\rho_3 \mathbf{b}\mathbf{1}^{\sf T}-\mathbf{1}\mathbf{y}_3^{\sf T}-\mathbf{y}_4\mathbf{1}).
    \end{align}
    The solution to the above equation can be obtained by taking the Kronecker product and then solving an extended linear equation. 
    However, solving this extended linear equation requires computing the inverse of a matrix of size $N_cN_{slot}\times N_cN_{slot}$, which incurs a computational complexity of $\mathcal{O}(N_c^3N_{slot}^3)$ at each step, making this approach computationally prohibitive.
    To mitigate this computational burden, we rewrite the equation in \eqref{eq:sylvester_eqaution} in the following form:
    \begin{align}\label{eq:sylvester_eqaution2}
        \mathbf{AX+XB=C},
    \end{align} 
    where $\mathbf{A}=2\mathbf{G}+(\rho_1+\rho_2)\mathbf{I}+2\rho_3\mathbf{11}^{\sf T}$, $\mathbf{B} = 2\rho_3\mathbf{11}^{\sf T}$, and
    \begin{align*}
        \mathbf{C}&=(\rho_1\mathbf{Z}_1+\rho_2\mathbf{Z}_2-\mathbf{Y}_1-\mathbf{Y}_2) \nonumber \\
        &~~~+2(\rho_3N_{b}\mathbf{11}^{\sf T}+\rho_3 \mathbf{b}\mathbf{1}^{\sf T}-\mathbf{1}\mathbf{y}_3^{\sf T}+\mathbf{y}_4\mathbf{1}).
    \end{align*} 
    This equation is known as the Sylvester equation \cite{Sylvester_equation}.
    We then show that this equation has a unique solution, as given in the following proposition:
    \begin{prop} \label{prop:sylvester_uniqueness}
        The equation in \eqref{eq:sylvester_eqaution2} has a unique solution $\mathbf{X}$.
    \end{prop}
    \begin{IEEEproof}
    The Sylvester equation in \eqref{eq:sylvester_eqaution2} has a unique solution $\mathbf{X}$ if and only if $\mathbf{A}$ and $\mathbf{-B}$ do not share any eigenvalue \cite{Sylvester_equation}.
    As the matrix $\mathbf{A}$ in \eqref{eq:sylvester_eqaution2} is composed of two positive definite matrices $2\mathbf{G}$ and $(\rho_1+\rho_2)\mathbf{I}$ and one positive semidefinite matrix $\mathbf{2}\rho_3 \mathbf{1}\mathbf{1}^T$, $\mathbf{A}$ is positive definite matrix.
    On the other hand, $\mathbf{-B}$ is a negative semidefinite matrix. 
    Therefore, $\mathbf{A}$ and $\mathbf{-B}$ do not share any eigenvalue, and the Sylvester equation \eqref{eq:sylvester_eqaution2} has a unique solution.
    \end{IEEEproof}
    
    The Bartels-Stewart algorithm is a well-known numerical method for solving the Sylvester equation \cite{Bartels-Steward_algorithm}.
    It first applies the Schur decomposition using QR decomposition to transform $\mathbf{A}$ and $\mathbf{B}$ into quasi-triangular forms.
    Then, back-substitution is used to solve the quasi-triangular form of the Sylvester equation.
    The computational complexity of this algorithm is $\mathcal{O}(N_c^3+N_{slot}^3)$, which is significantly lower than that of the Kronecker-based approach. Finally, gradient ascent is applied to update the dual variables $(\mathbf{Y}_1, \mathbf{Y}_2, \mathbf{y}_3, \mathbf{y}_4)$. 
    The resulting $\ell_2$-box ADMM steps are summarized in Algorithm \ref{alg:l2ADMM}.

    \begin{algorithm}[t]
        \caption{$\ell_2$-box ADMM}\label{alg:l2ADMM}
        {\small \begin{algorithmic}[1]
            \REQUIRE $\mathbf{b}, \mathbf{G}, \rho_1, \rho_2, \rho_3, \text{ and }\gamma$
            \STATE  \textbf{Initialize:} $\xi_{\min} \gets 0, \xi_{\max}\gets 1$
            \REPEAT
            \STATE \!\! Obtain $\mathbf{Z}_1^{k+1}$ from \eqref{eq:l2ADMM_Z1}.
            \STATE \!\! Obtain $\mathbf{Z}_2^{k+1}$ from \eqref{eq:l2ADMM_Z2}.
            \STATE \!\! Obtain $\mathbf{X}^{k+1}$ by solving \eqref{eq:sylvester_eqaution} using the Bartels-Steward algorithm.
            \STATE \!\! $\mathbf{Y}_1^{k+1}\gets \mathbf{Y}_1^k+\gamma\rho_1(\mathbf{X}^{k+1}-\mathbf{Z}_1^{k+1})$.
            \STATE \!\! $\mathbf{Y}_2^{k+1}\gets \mathbf{Y}_2^k+\gamma\rho_2(\mathbf{X}^{k+1}-\mathbf{Z}_2^{k+1})$.
            \STATE \!\! $\mathbf{y}_3^{k+1}\gets \mathbf{y}_3^k+\gamma\rho_3(\{\mathbf{X}^{k+1}\}^{\sf T}\mathbf{1}-N_b\cdot \mathbf{1})$.
            \STATE \!\! $\mathbf{y}_4^{k+1}\gets \mathbf{y}_4^k+\gamma\rho_3(\mathbf{X}^{k+1}\mathbf{1}-\mathbf{b})$.
            \UNTIL
            \STATE $\mathbf{X}_{out}\gets \mathrm{round}(\mathbf{X})$.
            \ENSURE $\mathbf{X}_{out}$
        \end{algorithmic}}
    \end{algorithm}
    
    \subsection{Proposed Alternative Optimization Algorithm}
    We propose an AO algorithm to solve the problem $\mathcal{P}_1$ based on the bisection method and the ADMM algorithms.
    The overall procedure of the proposed AO algorithm is summarized in  Algorithm \ref{alg:Alternative Optimization}.
    In our algorithm, we initially set  $x_i^t=\frac{b_i}{N_{slot}}$, ensuring that the constraint $b_i=\sum_{t=1}^{N_{slot}} x_i^t,~ \forall i\in\mathcal{N}_c$ is satisfied.
    Under this initialization, the decoding success probability becomes
    \begin{align}\label{eq:P_di_inital}
        P_{d,i}^{low}(\mathbf{X}) =1-\sum_{j\neq i}\frac{ g_{ij} N_j \alpha_j}{N_{slot}N_R(\frac{g_{ii}}{\gamma_{th}}-\rho^{-1})}.
    \end{align}
    After the initialization, the bisection method in Algorithm \ref{alg:Bisection} is employed to obtain $\mathbf{b}^{(k)}$ while keeping the beam-hopping pattern $\mathbf{X}^{(k)}$ fixed.
    Then, the ADMM algorithm (Algorithm \ref{alg:ADMM}) or $\ell_2$-box ADMM algorithm (Algorithm \ref{alg:l2ADMM}) is employed to determine the beam-hopping pattern $\mathbf{X}^{(k+1)}$ based on  $\mathbf{b}^{(k)}$.
    Since the ADMM and $\ell_2$-box ADMM algorithms treat $\mathbf{X}$ as a continuous-valued matrix and apply a rounding operation at the end, some constraints may not be satisfied.
    To satisfy the constraints, in Steps 5--15, we adopt a simple greedy mechanism as post-processing that removes excess beam assignments or adds insufficient assignments that maximize $P_{suc,i}^{low}$ for every time slot.

   
    \begin{algorithm}[t]
        \caption{Proposed Alternative Optimization Algorithm}\label{alg:Alternative Optimization}
        {\small \begin{algorithmic}[1]
            \REQUIRE $N_{AO}$
            \STATE  \textbf{Initialize:} $k=1$, $x_i^t=\frac{b_i}{N_{slot}}$, $P_{d,i}(\mathbf{X})$ is computed from \eqref{eq:P_di_inital}.
            \REPEAT
            \STATE Obtain $\mathbf{b}^{(k)}$ with fixed $\mathbf{X}^{(k)}$ by using Algorithm \ref{alg:Bisection}.
            \STATE Obtain $\mathbf{X}^{(k+1)}$ with fixed $\mathbf{b}^{(k)}$ by using Algorithm \ref{alg:ADMM} or \ref{alg:l2ADMM}.
            \REPEAT
            \STATE Evaluate $P_{suc,i}^{low}(\mathbf{X}, \mathbf{b})$ with beam-hoping pattern $\mathbf{X}$.
            \STATE Set $x^t = \sum_{i\in\mathcal{N}_c} x_i^t$.
            \IF {$x^t < N_b$}
            \STATE \!\!Assign $x_i^t=1$ to the $N_b-x^t$ cells with the lowest $P_{suc,i}^{low}$.
            \ENDIF
            \IF {$x^t > N_b$}
            \STATE \!\!Assign $x_i^t=0$ to the $x^t-N_b$ cells with the highest $P_{suc,i}^{low}$. 
            \ENDIF
            \STATE $t \leftarrow t+1$.
            \UNTIL $t \leq N_{slot}$
            \STATE $k \leftarrow k+1$.
            \UNTIL $k \leq N_{AO}$
            \ENSURE $\mathbf{X}^{(k)}$
        \end{algorithmic}}
    \end{algorithm}

    \subsection{Computational Complexity}
    \textcolor{black}{The computational complexity of the proposed AO framework in Algorithm~\ref{alg:Alternative Optimization} is analyzed on a per outer iteration basis, consisting of two stages: the bisection method and ADMM-based optimization.
    The bisection method (Algorithm~\ref{alg:Bisection}) has a complexity that includes one-time initialization of $P_{d,i}^{low}(\mathbf{X})$ with cost $\mathcal{O}(N_c^2 N_{slot})$ and $I_{B}$ bisection iterations, each requiring $\mathcal{O}(N_c)$ operations, yielding a total complexity of $\mathcal{O}(N_c^2 N_{slot} + I_{B} N_c)$.
    For ADMM methods, the standard ADMM (Algorithm~\ref{alg:ADMM}) performs $I_{A}$ iterations of matrix-matrix multiplications with complexity $\mathcal{O}(I_{A} N_c^2 N_{slot})$, while the $\ell_2$-box ADMM (Algorithm~\ref{alg:l2ADMM}) solves Sylvester equations using the Bartels-Stewart algorithm with complexity $\mathcal{O}(I_{L2A} (N_c^3 + N_{slot}^3))$.
    The total complexity over $I_{AO}$ outer loops is $\mathcal{O}(I_{AO}(I_B N_c + I_A N_c^2 N_{slot}))$ for the standard ADMM and $\mathcal{O}(I_{AO} (I_B N_c + I_{L2A} (N_c^3 + N_{slot}^3)))$ for $\ell_2$-box ADMM.}

	\section{Simulation Results}\label{Sec:Simulation}
    
    \begin{table}[t]
        \centering
        \caption{Simulation Parameters}
        \label{tab:simulation_parameters}
        \begin{tabular}{c|c}
            \hline
            \rowcolor{gray!10}
            \textbf{Parameter} & \textbf{Value} \\ \hline
            Rx $G/T$ & 1.1 dB/K \\ 
            Antenna aperture & 2m \\ 
            Beam radiation pattern $G_r(\theta)$ & Provided in \cite{3GPP-NTN_study}
            \tablefootnote{\textcolor{black}{
            The beam radiation pattern is based on the reflector antenna model from 3GPP NTN study (TR 38.811) \cite{3GPP-NTN_study}. Nevertheless, the proposed frameworks are not dependent on specific antenna models and can be broadly applied to various antenna types such as phased array antennas.}} \\ 
            UE transmit power $P_{tx}$ & 200 mW (23 dBm) \\ 
            Bandwidth per RB $B$ & 1 MHz/RB \\ 
            Frequency band $f_c$ & 2 GHz \\ 
            Satellite height $h$ & 600 km \\ 
            Boltmann constant $k$ & -228.6 dBw/K/Hz \\ 
            Number of cells $N_c$ & 80 \\ 
            Number of beams $N_b$ & 6 \\ 
            Number of time slots $N_{slot}$ & 64 \\ 
            Number of resource blocks $N_{R}$ & 20 \\ 
            Average number of users $N_{avg}$ & 1000 \\ 
            Population scaling factor $\beta$ & 0.5 \\ 
            Weight of two distinct scenarios $\eta$ & 0.3 \\ 
            Device activation probability $\alpha$ & 0.01 \\ 
            Bisection iterations  & 100 \\ 
            ADMM iterations $I_{A}$ & 300 \\ 
            ADMM update rate $\gamma$  & 1 \\ 
            AO iterations in Algorithm \ref{alg:Alternative Optimization} $I_{AO}$ & 5 \\
            SINR threshold $\gamma_{th}$ & 5 dB \\ \hline
        \end{tabular}
        \vspace{-3mm}
    \end{table}
    
    This section evaluates the performance of the proposed beam-hopping pattern design algorithms through simulations. 
    Traffic demand follows the proposed traffic demand model described in Sec.~\ref{Sec:Traffic_demand_model}.
    The average number of devices per cell and the population scaling factor are set to $N_{avg}=1000$ and  $\beta=0.5$, respectively.
    For simplicity, all devices within the coverage area are assumed to have an identical activation probability $\alpha_i=0.01, ~ \forall i\in\mathcal{N}_c$.
    Satellite system parameters are set according to 3GPP TR 38.811 \cite{3GPP-NTN_study}.
    \textcolor{black}{The satellite orbits at 600 km altitude and serves $N_c=80$ cells.
    These cells are the closest to the satellite's nadir, forming a concentrated service area.}
    The satellite has multi-beam antennas that can generate up to $N_b=6$ beams, with beam-hopping windows divided into $N_{slot}=64$ time slots.
    Ground devices select one of the $N_{R}=20$ resource blocks for transmission.
    The carrier frequency is $f_c=2 \text{GHz}$ and the SINR threshold is $\gamma_{th}=5\mathrm{dB}$. 
    Performance evaluation uses 10,000 randomly sampled satellite positions.
    The weight parameter $\eta$, which balances the population-based and uniform traffic distributions, is set to 0.3, creating significant traffic demand asymmetry.
    For ADMM, $\rho_2$ is set to 2.2 times of $\rho_1$ while $\ell_2$-box ADMM uses identical values for $\rho_1$, $\rho_2$, and $\rho_3$.
    To accelerate convergence, parameters $\rho_i$ are increased by a factor of 1.01 each iteration until reaching 3.6, starting from small initial values determined by traffic demand disparity.
    Simulation parameters are summarized in Table \ref{tab:simulation_parameters}. 
    For performance comparison, we consider the following beam-hopping methods: 
    
    \begin{enumerate}[label=(\roman*)]
        \item \emph{Random selection}:  Randomly selects $N_b$ cells from the total $N_c$ cells with replacement at each time slot.
        
        \item \emph{Round robin}: Follows a periodic scheduling strategy, where the satellite sequentially serves the covered cells in a cyclic order.
        Each cell is selected once every $N_c/N_b \approx 13.33$ time slots.
        \textcolor{black}{The computational complexity of this method is $\mathcal{O}(N_{slot} N_b)$.}
        
        \item \emph{Greedy}: Selects $N_b$ cells with the lowest ratio of devices to allocated beams at each time slot.
        \textcolor{black}{The computational complexity of this method is $\mathcal{O}(N_{slot}^2 N_c)$.}
        
        \item \textcolor{black}{\emph{Genetic}: The genetic method \cite{Beam-hopping_genetic} employs selection, crossover, and mutation to evolve solutions $\mathbf{X}$. 
        The population size is set to $P = 100$ and the number of generations is set to $I_{G} = 250$}.
        \textcolor{black}{The computational complexity of this method is $\mathcal{O}(I_{G}PN_{slot}^2 N_c)$, where $I_{G}$ denotes the number of generations and $P$ represents the population size.}
        
        \item \emph{Bisection + LP relaxation (B-LP)}: The B-LP method adopts the same AO framework, but alternates between the bisection method and linear programming (LP) relaxation. This addresses the problem $\mathcal{P}_3^\prime$ by substituting the binary constraint with a continuous interval and employing the commercial solver MOSEK \cite{MOSEK}.
        \textcolor{black}{The computational complexity of this method is $\mathcal{O}(I_{AO} (I_B N_c + I_{int}(N_{slot}N_c)^{3.5})$, where $I_{int}$ denotes the number of interior method iterations.}
        
        \item \emph{Proposed B-A/B-L2A}: The proposed method from Algorithm~\ref{alg:Alternative Optimization}, combining the bisection method with ADMM in Algorithm~\ref{alg:ADMM} (Proposed B-A) or with $\ell_2$-box ADMM in Algorithm~\ref{alg:l2ADMM} (Proposed B-L2A).
    \end{enumerate}

    \subsection{Success Transmission Probability}
    \begin{figure}[t]
    	\centering
    	{\epsfig{file=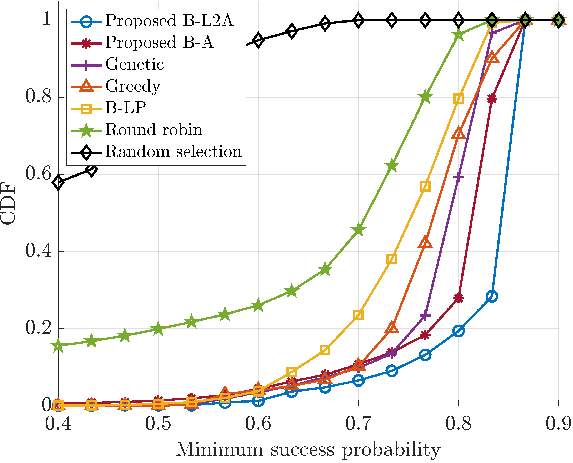, width=6.8cm}}
        \caption{Comparison of the CDFs of the minimum success transmission probability for various beam-hopping methods.}
    	\label{fig:cdf_03_global}
        \vspace{-1mm}
    \end{figure}

    Fig.~\ref{fig:cdf_03_global} compares the cumulative distribution functions (CDFs) of the minimum success transmission probability for various beam-hopping methods using randomly selected satellite locations across the globe.
    The results show that the proposed methods achieve a minimum success transmission probability of 0.6 for the worst performing cell at most satellite locations and consistently outperform other benchmarks.
    \textcolor{black}{At the 30th percentile, the proposed B-A and B-L2A methods surpass all benchmarks by more than 3.5\% and 6.8\%, respectively.
    While the genetic and greedy methods show competitive results, they fail to guarantee a minimum success transmission probability of 0.8 for 60\% of the locations.}
    When comparing the two proposed methods, we observe that the B-L2A method consistently demonstrates superior performance, whereas the B-A method shows noticeable degraded performance in the low minimum success probability region.
    This discrepancy arises because the original ADMM fails to satisfy the beam capacity constraint $C_3$ of problem $\mathcal{P}^\prime_3$ under extreme traffic disparity.
    Specifically, the rounding operation in ADMM suppresses allocations to low-traffic demand cells to zero, leading to significant performance deterioration.
    These results confirm that our $\ell_2$-box relaxation in Sec.~\ref{Sec:l2-ADMM} improves the stability of the beam-hopping pattern determination process.
    
	\begin{figure}[t]
    	\centering
    	{\epsfig{file=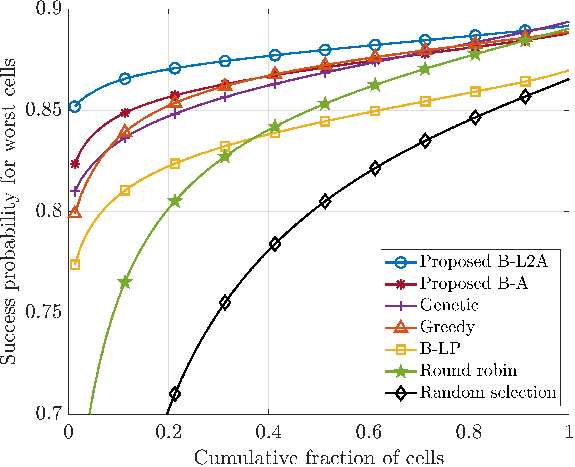, width=6.8cm} 
        \caption{Comparison of the success transmission probabilities across the cumulative fraction of worst-performing cells for various beam-hopping methods.}
    	\label{fig:cumulative_fraction_03_global}}
        \vspace{-2mm}
    \end{figure}
    
    Fig.~\ref{fig:cumulative_fraction_03_global} compares the success transmission probabilities as the cumulative fraction of worst-performing cells increases from 0 to 1.
    As the fraction increases, the plotted probability transitions from the minimum success transmission probability (i.e., $\min_{i\in\mathcal{N}_c} P_{suc,i}({\bf X})$) to the average success transmission probability (i.e., $\frac{1}{|\mathcal{N}_c|}\sum_{i\in\mathcal{N}_c} P_{suc,i}({\bf X})$).
    Fig.~\ref{fig:cumulative_fraction_03_global} shows that both the proposed B-A and B-L2A methods outperform other benchmark methods when the fraction is below 0.3, demonstrating their superiority in improving the success transmission probability of the worst-performing cells.
    Moreover, the proposed B-A and B-L2A methods exhibit comparable performance at a fraction of 1 (i.e., the average success transmission probability), even though maximizing the average success transmission probability is not the primary objective of our optimization problem.
    These results confirm that our methods effectively balance overall performance and fairness in beam-hopping pattern design.
    \textcolor{black}{The round robin, greedy, and genetic methods achieve comparable performance to the proposed methods at a fraction of 1, but they suffer from significant performance degradation as the fraction decreases.}
    \textcolor{black}{The B-LP method, despite employing the same AO framework as the proposed methods, exhibits a significant performance gap compared to our methods due to the large integrality gap, leading to inefficient solutions after rounding.
    This result validates the effectiveness of our ADMM-based approaches in solving the problem $\mathcal{P}_3^\prime$ by enforcing the binary constraint with an explicit update step.}

    \subsection{Visualization of Traffic Demand and Beam-Hppping Pattern} 
    \begin{figure}[t]
    \centering
    \begin{subfigure}[Minimum sucess probability = $0.8772$]
    	{\epsfig{file=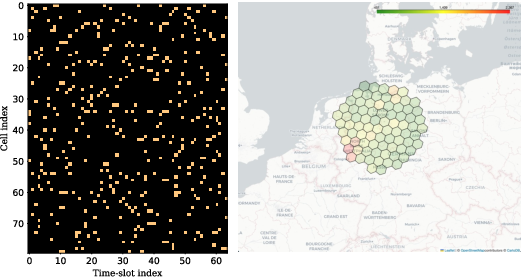, width=0.4\textwidth}}
    \end{subfigure}
    \begin{subfigure}[Minimum sucess probability = $0.7815$]
    	{\epsfig{file=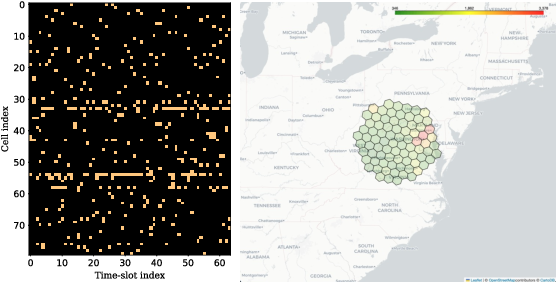, width=0.4\textwidth}}
    \end{subfigure}
    \caption{Visualization of our population-based traffic demand model alongside the corresponding beam-hopping patterns generated by the proposed B-L2A method for different scenarios. 
    Left: optimized binary beam-hopping patterns $\mathbf{X}$.
    Right: relative traffic demand distributions across cells.}
    \label{fig:beam_hopping_pattern_examples}
    \vspace{-2mm}
    \end{figure}
  
    Fig.~\ref{fig:beam_hopping_pattern_examples} visualizes our population-based traffic demand model alongside the corresponding beam-hopping patterns generated by the proposed B-L2A method for different scenarios.
    Each subplot's left side shows the beam-hopping pattern $\mathbf{X}$ as a 0-1 heatmap, where colored regions indicate $x_i^t=1$, representing an active beam assigned to cell $i$ at time slot $t$, while black regions represent $x_i^t=0$.
    Each subplot's right side depicts the $N_c$ cells served by the satellite on a global map, where the color intensity represents the relative traffic demand of each cell. 
   
    Fig.~\ref{fig:beam_hopping_pattern_examples}(a) shows a scenario with relatively low demand contrast, where the ratio between the highest and lowest traffic demand is about 5.
    In this case, beam allocation is more evenly distributed, with the maximum number of allocated beams being only 13.
    In contrast, Fig.~\ref{fig:beam_hopping_pattern_examples}(b) illustrates a scenario with high traffic demand imbalance, where the highest cell traffic demand is 10 times higher than the minimum demand. 
    High traffic demand concentrates in a few urban cells near Washington DC, while other cells have low traffic demand.
    To handle this imbalance, the proposed B-L2A method allocates large amounts of beams on busy cells, while only 2-3 beams are assigned to low-demand cells.
    The beam-hopping pattern allocates a maximum of 37 beams, which is more than twice that of the previous scenario. 
    This is because high-demand cells are located nearby, inevitably causing increased inter-beam interference.
    These results demonstrate that the proposed B-L2A method effectively generates beam-hopping patterns tailored to traffic demand variations, adapting to diverse traffic demand scenarios.
    This figure also shows that scenarios with high traffic demand disparity exhibit relatively lower minimum success transmission probabilities, whereas more balanced scenarios achieve higher performance. 
    \textcolor{black}{This indicates that beam-hopping has greater difficulty in achieving high success transmission probability in highly imbalanced traffic demand scenarios than in uniform traffic demand scenarios.}

   \begin{figure}[t]
    	\centering
    	{\epsfig{file=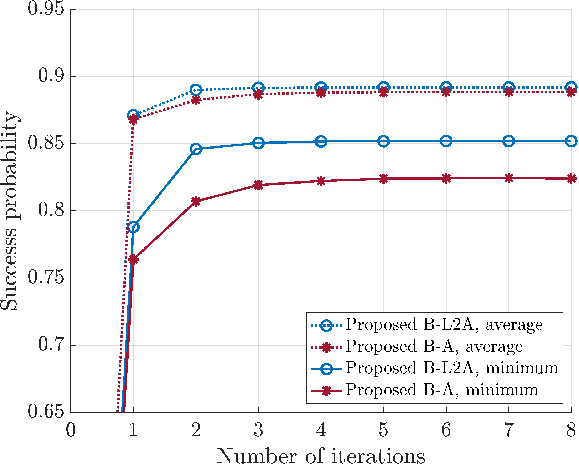, width=6.8cm}} 
        \caption{Convergence curves of the proposed B-A and B-L2A methods versus AO iterations.} \vspace{-3mm}
    	\label{fig:convergence_AO}
    \end{figure}

    \subsection{Convergence of Alternative Optimization}
    Fig.~\ref{fig:convergence_AO} illustrates the convergence curves of the proposed methods versus the number of AO iterations.
    Both B-A and B-L2A demonstrate a steady increase in success probability over approximately five iterations, followed by stable convergence. 
    \textcolor{black}{While the proposed methods maximize the average success transmission probability during the decoding success probability maximization phase, they also improve the minimum success transmission probability throughout the AO iterations.}
    These results confirm the reliability and effectiveness of the AO approach used in the proposed methods.
    
    \subsection{Scalability and Computational Complexity}
    
    \begin{figure}[t]
    	\centering
    	{\epsfig{file=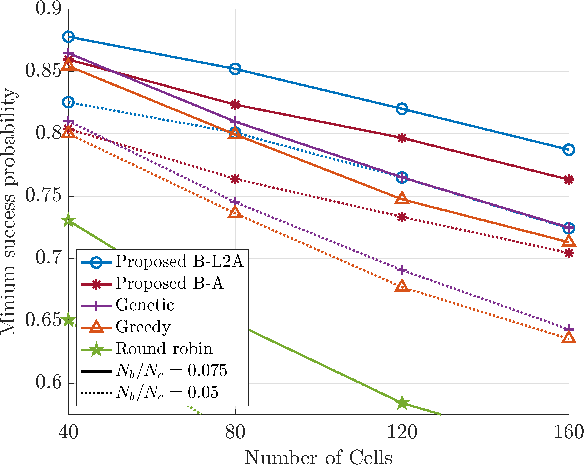, width=6.8cm}} \vspace{-1mm}
        \caption{Comparison of the minimum success transmission probabilities in terms of the number of cells and $N_b/N_c$ ratios.} \vspace{-3mm}
    	\label{fig:scalability}
    \end{figure}
    \begin{table}[t]
        \centering
        \caption{Computational time (in seconds) required by each algorithm to generate a beam-hopping pattern.}
        \label{tab:computation_complexity}
        \begin{tabular}{|c|c|c|c|c|c|}
            \hline
            \rowcolor{gray!10}
            \textbf{Cells}& \textbf{Round robin} & \textbf{Greedy}& \textbf{Genetic} & \textbf{B-A} & \textbf{B-L2A} \\ 
            \hline
            40 & $1.0\times10^{-4}$ & $6.6\times10^{-4}$ & 36.72 &  0.1453 & 0.9636 \\
            80 & $1.0\times10^{-4}$ & $7.8\times10^{-4}$ & 88.95 & 0.2621 & 2.5573 \\
            120 & $1.1\times10^{-4}$ & $8.9\times10^{-4}$ & 186.26 & 0.4142 & 5.2156\\
            160 & $1.0\times10^{-4}$ & $1.1\times10^{-3}$ & 455.92 & 0.5986 & 7.6184 \\
            \hline
        \end{tabular}
        \vspace{-3mm}
    \end{table}
    \textcolor{black}{
    Fig.~\ref{fig:scalability} compares the minimum success transmission probability as a function of the total number of served cells ($N_c$), which varies from 40 to 160.
    To maintain a consistent resource ratio across these scenarios, the number of available beams ($N_b$) is scaled accordingly, maintaining fixed $N_b/N_c$ ratios of 0.075 and 0.05.
    Fig.~\ref{fig:scalability} shows that the minimum success probability for all methods degrades as the number of cells increases, when the ratio $N_b/N_c$ is fixed.
    This occurs because serving a larger number of cells requires covering a wider geographical area, which inevitably includes cells at lower elevation angles.
    These cells experience higher pathloss and reduced antenna gain, leading to lower overall minimum performance.
    The results also confirm that a higher $N_b/N_c$ ratio improves performance, as the capacity gains from additional beam availability exceed the performance penalty from increased inter-beam interference.
    When comparing the algorithms, our proposed B-L2A and B-A methods consistently outperform the greedy and round robin benchmarks across all tested scenarios.
    While the genetic method performs reasonably well for a small number of cells, and even shows marginal gains over B-A for small-scale ($N_c = 40$), its performance degrades much more sharply than our proposed methods as $N_c$ increases.
    This performance gap becomes significantly pronounced at lower $N_b/N_c$ ratios, where resource allocation becomes more critical.
    } 

    \textcolor{black}{
    Table~\ref{tab:computation_complexity} presents the computational time (in seconds) required by each algorithm to generate a beam-hopping pattern as $N_c$, the key parameter affecting complexity, is varied.
    The experiments were performed using an Intel Core i9-12900K processor with Ubuntu 22.04 LTS.
    The table shows that the proposed methods have a clear trade-off between performance and computational complexity.
    The B-L2A algorithm consistently achieves the highest success transmission probability across all tested configurations.
    This superior performance, however, comes at the cost of higher computational complexity.
    In contrast, the B-A algorithm provides competitive performance, but with more moderate computational complexity.
    Therefore, the choice between the algorithms depends on system priorities: B-L2A provides the highest performance, whereas B-A offers a computationally less demanding yet highly effective alternative. Although the genetic method demonstrates performance comparable to the proposed B-A method in certain scenarios (see Fig.~\ref{fig:scalability}), it incurs significantly higher computational complexity, making it impractical compared to the proposed approaches.} 

	\section{Conclusion}
    In this paper, we addressed the beam-hopping pattern design problem for LEO satellite communication systems with grant-free random access.
    To tackle this challenging binary optimization problem, we proposed an AO-based algorithm that iteratively maximizes both the collision avoidance rate and the decoding success probability.
    Specifically, the collision avoidance rate is optimized by determining the number of beam illuminations for each cell using a bisection method.
    Given this allocation, the ADMM framework is employed to determine the beam-hopping pattern that maximizes the decoding success probability by mitigating inter-beam interference.
    Through extensive simulations, we validated the effectiveness of the proposed algorithms, demonstrating substantial performance gains over other benchmarks.
    Future research directions include extending our optimization framework by incorporating communication resource allocation in the frequency or power domain.
    Another promising avenue is the joint optimization of beam-hopping patterns for multi-satellite systems.
    

    \appendices
    \section{Proof of Lemma~\ref{lem:bisection uniqueness}}\label{apdx:bisection uniqueness}
    We prove Lemma~\ref{lem:bisection uniqueness} by contradiction.
    The function $f_i(\xi)$ in \eqref{eq:Bisection b} satisfies the constraints $C_1^\prime,C_3 \text{ and } C_4$ in $\mathcal{P}_2^\prime$. For the feasible point $(\mathbf{b}, \xi)$, we have $_i(\xi)\leq \max(\lceil \bar{f}_i(\xi) \rceil, 1) \leq b_i$, $\forall i \in \mathcal{N}_c$.
    Suppose that $\sum_i b_i^*<N_{slot}N_b$, then $\sum_{i} f(\xi^*)\leq \sum_i b_i^* \leq N_{slot}N_b - 1$.
    From the condition of the lemma,
    \begin{align}
        \lim_{\xi\rightarrow \xi^* + } \sum_i f_i(\xi) - 1\leq \sum_i f(\xi^*) \leq N_{slot}N_b -1.
    \end{align}
    From the inequality, we can find that there exists a constant $\epsilon > 0$ which satisfies $\sum_i f_i (\xi^* +\epsilon) \leq N_{slot}N_b$. 
    This implies that we can find a feasible $\xi$ larger than $\xi^*$, which contradicts the optimality of $\xi^*$.
    Therefore, $\sum_i b_i^*= N_{slot}N_b$ holds.

    We again exploit the contradiction to show that the optimal point $(\mathbf{b}^*, \xi^*)$ is unique.
    Let $\xi^*$ be an optimal value of the problem $\mathcal{P}_2^\prime$. 
    Since the objective function is $\xi$ itself, the optimal value $\xi^*$ is unique.
    On the other hand, any $\mathbf{b}$ which satisfies all constraints of the problem can be an optimal value and the function $f_i(\xi^*)$ is a candidate for $\mathbf{b}$.
    We set $b_i^*=f_i(\xi^*), ~\forall i\in\mathcal{N}_c$ with ${\mathbf{b}}^* = [{b}_1^*,\cdots, {b}_{N_c}^*]^{\sf T}$. This allocation is an optimal point as $f_i(\xi)$ satisfies all the constraints.
    Suppose that there exists another optimal point $(\bar{\mathbf{b}}^*, \xi^*)$ such that $\bar{\mathbf{b}}^* \neq \mathbf{b}^*$, where $\bar{\mathbf{b}}^* = [\bar{b}_1^*,\cdots, \bar{b}_{N_c}^*]^{\sf T}$.
    There are some indexes $i\in\mathcal{N}_c$ where $\bar{b}_i^*\neq b_i^*$. 
    Both ${\mathbf{b}}^*$ and $\bar{\mathbf{b}}^*$ satisfy the equation $\sum_i b_i^* = \sum_i \bar{b}_i^*= N_{slot}N_b$.
    Therefore, there exists $i\in\mathcal{N}_c$ such that $\bar{b}_i^* < b_i^* = f_i (\xi^*)$, which contradicts $f_i (\xi^*)\leq \bar{b}_i^*$.
    
    \section{Proof of Theorem~\ref{lem:bisection convergence}}\label{apdx:bisection convergence}
    The function $f_i(\xi)$ in \eqref{eq:Bisection b} increases monotonically for $0<\xi<P^{low}_{d,i}(\mathbf{X})$. Therefore, the sum $\sum_if_i(\xi)$ also increases monotonically for $0<\xi<P^{low}_{d,i}(\mathbf{X})$.
    Suppose that $\xi_\ell^k\leq \xi^* \leq \xi_u^k$ and the optimal point $(\mathbf{b}^*, \xi^*)$, which is unique, satisfies $\sum_{i\in\mathcal{N}_c}b_i^*= N_{slot}N_b$.
    If $\sum f_i(\xi_m) \leq N_{slot} N_b$, $\xi_m^k \leq \xi^*$ due to the monotonically increasing property, and we can set as lower bound $\xi_\ell^{k+1}=\xi_m^k$.
    If $\sum f_i(\xi_m) > N_{slot} N_b$, $\xi_m^k > \xi^*$ so that we can set as upper bound $\xi_u^{k+1}=\xi_m^k$.
    Therefore, if $\xi_u^k$ and $\xi_\ell^k$ are the upper and lower bound at iteration $k$, the $\xi_u^{k+1}$ and $\xi_\ell^{k+1}$ are also the upper bound and lower bound at iteration $k+1$, with one updated to $\xi_m^k$.
    The error $\epsilon^k$ can expressed as
    \begin{align}
        \vert\epsilon^{k+1}\vert=\vert\xi^* - \xi_\ell^{k+1}\vert \leq \xi_u^{k+1} - \xi_\ell^{k+1} = \frac{1}{2} (\xi_u^{k} - \xi_\ell^{k}) = \frac{1}{2} \epsilon^{k}.
    \end{align}
    This shows linear convergence of the bisection method to the unique optimal point.


    \bibliographystyle{IEEEtran}
    \bibliography{Reference}

\end{document}